\documentclass{PoS}
\usepackage{bm}
\usepackage{amsmath}
\usepackage{amsfonts}
\usepackage{amssymb}

\title{Recent developments in nuclear structure theory:\\ an outlook on the muonic atom program}

\ShortTitle{Nuclear structure muonic atoms program}

\author{Oscar Javier Hernandez\\
        TRIUMF, 4004 Wesbrook Mall, Vancouver, BC, V6T 2A3, Canada and \\
       University of British Columbia, Vancouver, BC, V6T 1Z4, Canada\\
        E-mail: \email{javierh@phas.ubc.ca}}

\author{\speaker{Sonia Bacca}\\ %\thanks{A footnote may follow.}\\
        TRIUMF, 4004 Wesbrook Mall, Vancouver, BC, V6T 2A3, Canada\\
        University of Manitoba, Winnipeg, MB, R3T 2N2, Canada\\
        E-mail: \email{bacca@triumf.ca}}

\author{Kyle Andrew Wendt\\
  Lawrence Livermore National Laboratory, P.O. Box 808, L-414,\\ Livermore, California 94551, USA\\
E-mail: \email{wendt6@llnl.gov}}

\abstract{The discovery of the proton-radius puzzle and the subsequent deuteron-radius puzzle is fueling an on-going debate on possible explanations for the difference in the observed radii obtained from muonic atoms and from electron-nucleus systems.  
 Atomic nuclei have a complex internal structure that must be taken into account when analyzing experimental spectroscopic results. {\it Ab initio} nuclear structure theory provided the so far most precise estimates of important corrections to the Lamb shift in muonic atoms and is well poised to also investigate nuclear structure corrections to the hyperfine splitting in muonic atoms. Independently on  whether the puzzle is due to beyond-the-standard-model physics or not, nuclear structure corrections are a  necessary theoretical input to any experimental extraction of electric and magnetic radii from precise muonic atom measurements.
 
Here, we review the status of the calculations performed by the TRIUMF-Hebrew University group, focusing on the deuteron, and discuss preliminary results on  magnetic sum rules  calculated with two-body currents at next-to-leading order. Two-body currents will be an important ingredient in future calculations of nuclear structure corrections to the hyperfine splitting in muonic atoms.}

\FullConference{55th International Winter Meeting on Nuclear Physics\\
		23-27 January, 2017\\
		Bormio, Italy}

\begin{document}

\section{Introduction}

 In 2010, the CREMA (Charge Radius Experiment with Muonic Atoms) collaboration extracted the proton charge radius from measurements of the $2S-2P$ transition in muonic hydrogen ($\mu$H), a proton orbited by a muon. It was found  to deviate by about
7$\sigma$~\cite{Pohl:2010zza, Antognini:1900ns} with respect to the value obtained in decades of experiments  on both hydrogen spectroscopy  and  electron scattering off the proton.  This large discrepancy hinted towards new physics and created a lot of excitement in the  community. Interpretations of the discrepancy are being sought into systematic experimental errors,  novel aspects of hadronic structure, or  beyond-the-standard-model theories, leading to lepton universality violations.
% To date, no commonly accepted explanation  exists.
%If new forces are at play, it is interesting to ask the question of whether they couple differently to the proton or the neutron. 
 To investigate whether the discrepancy persists or changes with the nuclear mass number $A$ and proton number $Z$, the CREMA collaboration has embarked on a strong experimental program to extract the charge radii of light nuclei 
  by measuring the Lamb shifts in $\mu$-D, $\mu$-$^3$He$^+$ and $\mu$-$^4$He$^+$.

The Lamb shift is related to the charge radius $R_{c}$ by
\begin{equation}
\label{eq:LS}
\Delta E_{\rm LS} = \delta_{\rm QED}+\delta_{\rm FS}(R_{c})+\delta_{\rm TPE}.
\end{equation}
The three terms, from the largest to the smallest, are the quantum electrodynamics (QED) contributions,
% from vacuum polarization, lepton self energy, and relativistic recoil; 
the leading correction due to the finite size of the nucleus, $\delta_{\rm FS}(R_{c}) = \frac{m^{3}_{r}(Z\alpha)^{4}}{12}R^{2}_{c}$ (in $\hbar=c=1$ units and with $Z$ and $\alpha$ being the proton number and fine structure constant, respectively), and the two-photon exchange (TPE) contribution.

\begin{figure}[htb]
 \centerline{\includegraphics*[width=5.cm]{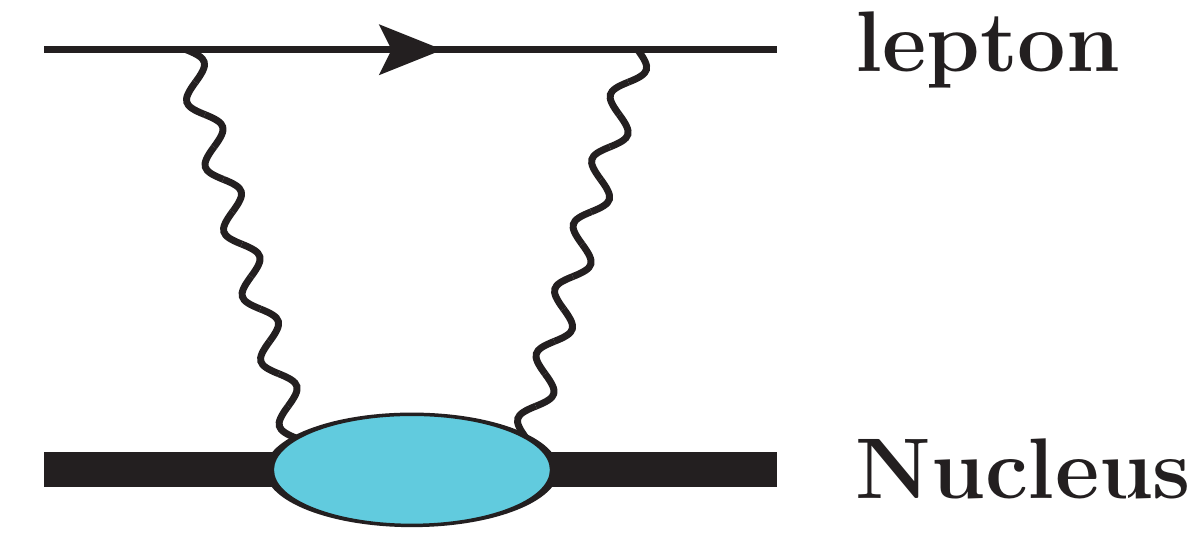}}
\caption{The lepton-nucleus two-photon-exchange. The blob denotes the excitation of the nucleus in the intermediate states between the two photons.}
\label{fig:tp}
\end{figure}

While quantum electrodynamical calculations of these atoms are extremely precise,  effects due to the structure of the nucleus constitute the main source of uncertainty and are the bottleneck to increase the precision of the extracted radius. Nuclear structure corrections appear via finite nuclear size effects -- precisely those effect that enable the extraction of the radius--, as well as via nuclear excitations in the TPE diagram. Here, virtual photons are exchanged between the lepton and the nucleus/hadron as shown in Fig.~\ref{fig:tp}.
The precision via which $\delta_{\rm TPE}$  can be calculated determines the precision in the extracted  charge radius. Independently on whether the puzzle is due to beyond-the-standard-model physics or not, precise calculations of nuclear structure corrections will always be needed and must accompany the experimental program aimed at extracting radii. 

\begin{table}[htb]
\centering
\caption{Experimental error bar in the measured Lamb-shift energy of muonic atoms $\delta_{\rm exp}(\Delta E_{\rm LS})$ 
 compared to the error bar in the theoretical calculation of the TPE  energy corrections to the Lamb-shift $\delta_{\rm th}(\Delta E_{\rm LS})$. Data taken from Ref.~\cite{Antognini:1900ns,science2016,Krauth_paper,Franke}.
}
\label{tab:1}
\begin{tabular}{l|l|l}
\hline\noalign{\smallskip}
&$\delta_{\rm exp}(\Delta E_{\rm LS})$& $\delta_{\rm th}(\Delta E_{\rm LS})$\\
\noalign{\smallskip}\hline\noalign{\smallskip}
$\mu$-H &2.3 $\mu$eV & 2 $\mu$eV \\
$\mu$-D                  & 3.4 $\mu$eV & 20 $\mu$eV \\
$\mu$-$ ^{3} {\rm He}^{+}$ & 0.08 meV  &  0.52 meV\\
\noalign{\smallskip}\hline
\end{tabular}
\end{table}

To appreciate the importance of nuclear structure corrections in nuclei, it is interesting to look at the experimental error bar via which the Lamb shift energy can be measured, $\delta_{\rm exp}(\Delta E_{\rm LS})$, and compare it to the theoretical error bar in the TPE calculations, $\delta_{\rm th}(\Delta E_{\rm LS})$. As shown in Table~\ref{tab:1}, while for the $\mu$-H case both errors are of the same order of magnitude, for $\mu$-D and $\mu-^3$He$^+$ the ratio between $\delta_{\rm th}(\Delta E_{\rm LS})$ and $\delta_{\rm exp}(\Delta E_{\rm LS})$ is about 6. This indicates, that for light muonic atoms TPE corrections constitute the real bottleneck to exploit the experimental precision in the extraction of the charge radius. 

So far, the TRIUMF - Hebrew University group has provided  the most precise determination of $\delta_{\rm TPE}$ corrections to the Lamb shift for $\mu$-D~\cite{Hernandez2014}, $\mu$-$^3$He$^+$ and $\mu$-$^3H$~\cite{Nevo16}, and  $\mu$-$^4$He$^+$~\cite{Ji2013,Nevo2014} using chiral effective field theory~\cite{entem2003,Epelbaum09} and phenomenological potentials~\cite{AV18} combined with state-of-the-art few-body calculational tools.

Contributions to $\delta_{\rm TPE}$ can be divided into the elastic Zemach term and the inelastic polarization term, i.e., $\delta_{\rm TPE} = \delta_{\rm Zem}+\delta_{\rm pol}$. Both can be further separated into nuclear $(\delta^{A})$ and nucleonic $(\delta^{N})$ components,  i.e., $\delta_{\rm TPE} = \delta^{A}_{\rm Zem}+\delta^{A}_{\rm pol}+\delta^{N}_{\rm Zem}+\delta^{N}_{\rm pol}$. 
 The inelastic nuclear term is called polarization term, since it is related to the polarizability of the nucleus,  i.e., the excitations of the nucleus over all its continuum spectrum due to the virtual absorption and subsequent emission of  photons, expressed by the blob in Fig.~\ref{fig:tp}.  Below we report our results for the various nuclei, also shown in Ref.~\cite{Javier2016}.
The uncertainty associated with each value is given in brackets and includes the numerical, nuclear model, and atomic physics  errors.  It is worth noticing that the uncertainties in $\delta_{\rm TPE}$ are slightly different than those shown in Table~\ref{tab:1}. This is due to the fact that the uncertainties in Table~\ref{tab:1} are taken from the analysis performed by colleagues~\cite{Krauth_paper,Franke} and do not include only our calculations, but an average with results of other groups as well~\cite{Pachucki11,Friar13,Carlsson14,Pachucki15}.

\begin{table}[htb]
\centering
\caption{Contributions to $\delta_{\rm TPE}$ of the Lamb shift in light muonic atoms, in meV, where we omit the proton-neutron subtraction term~\cite{Krauth_paper}.}
\label{tab:2}       % Give a unique label
%% For LaTeX tables use
\begin{tabular}{l|llll|l}
\hline\noalign{\smallskip}
 & $\delta^{A}_{\rm Zem}$ & $\delta^{A}_{\rm pol}$ & $\delta^{N}_{\rm Zem}$ & $\delta^{N}_{\rm pol}$ & $\delta_{\rm TPE}$  \\
\noalign{\smallskip}\hline\noalign{\smallskip}
$\mu$-D & -0.424(3) &  -1.245(19) & -0.030(2) & -0.028(2) & -1.727(20) \\
$\mu$-$ ^{3}{\rm H}$ & -0.227(6) & -0.473(17) & -0.033(2) & -0.034(16) &  -0.767(25)  \\
$\mu$-$ ^{3} {\rm He}^{+}$ & -10.49(24) &  -4.17(17)& -0.52(3)  & -0.28(12) & -15.46(39) \\
$\mu$-$ ^{4}{\rm He}^{+}$ & -6.29(28) & -2.36(14) & -0.54(3) & -0.38(22) &   -9.58(38) \\
\noalign{\smallskip}\hline
\end{tabular}
\end{table}

 In particular, here we want to concentrate on the deuteron, for which we have so far performed the most thorough calculations, by also analyzing the convergence of the chiral expansion, see Ref.~\cite{Hernandez2014}. Our results, together with others, have already been used by the CREMA collaborations to extract the value of the  charge radius from  muonic deuterium Lamb shift measurements~\cite{science2016}. Interestingly, in analogy to the proton case, such radius  revealed to be smaller, with about a $7\sigma$ deviation with respect to CODATA-2010 evaluations~\cite{Mohr:2012tt} and 3.5 $\sigma$ with respect  to spectroscopic extractions from ordinary deuterium alone~\cite{deut_spect}. Different from the proton case, in the so called ``deuteron-radius puzzle''  electron scattering data~\cite{Sick} are not precise enough to discriminate among muonic and electronic deuteron spectroscopy.

By using Eq.~(\ref{eq:LS}) one can also experimentally determine the size of  $\delta_{\rm TPE}$. Indeed, the $\mu$-D measurements in \cite{science2016} provided the left-hand-side of  Eq.~(\ref{eq:LS}), while 
the size of the deuteron can be determined from a combination of  measurements of isotope shift  in ordinary deuteron and muonic Lamb shift in the proton. Interestingly,  the measured $\delta_{\rm TPE}$ turns out to deviate 2.5$\sigma$~\cite{science2016} with respect to theoretical computations, including our work~\cite{Hernandez2014} and calculations by others, see e.g., Refs.~\cite{Pachucki11,Pachucki15}. While this fact certainly needs to be further investigated, with respect to the ~7$\sigma$ deviation between muonic deuteron and CODATA-2010 evaluation, this difference is minor. In the past we investigated the dependence of $\delta_{\rm TPE}$ on the nuclear potential used in input and found it to be small. 
\begin{figure}[htb]
\centering
 \includegraphics*[width=14.cm]{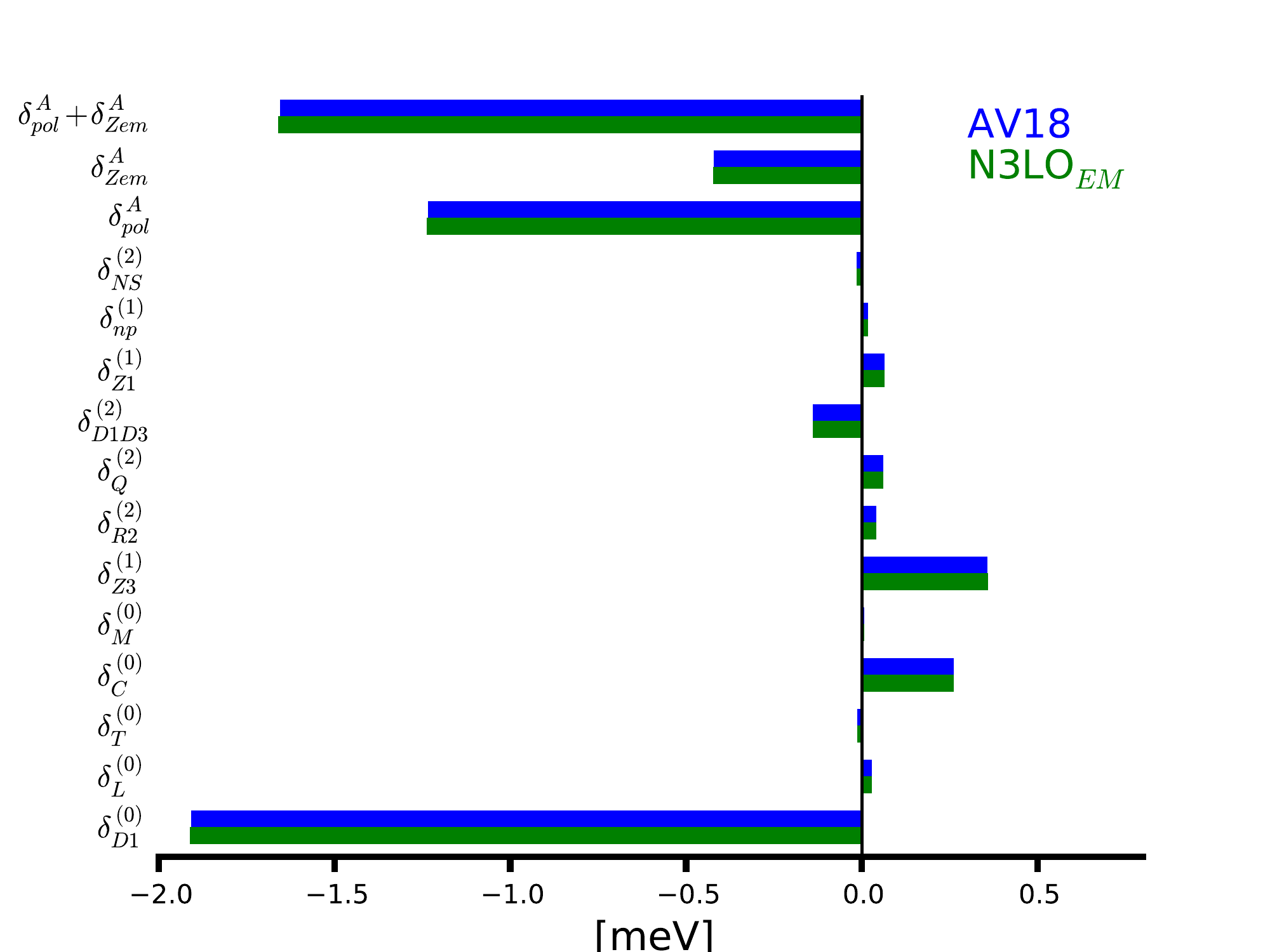}
\caption{Graphic representation of the various contributions 
		to $\delta^A_{pol} + \delta^A_{Zem}$ in the \mbox{2S-2P} Lamb shift 
		of $\mu$D, 
		calculated with the AV18~\cite{AV18} and a chiral effective field theory nuclear potential at N3LO~\cite{entem2003}.}
\label{fig:deut} 
\end{figure}
Below, we present a graphic representation of the various contributions to  $\delta^A_{pol} + \delta^A_{Zem}$  for the muonic deuteron case. This corresponds to $\delta_{\rm TPE}$, a part from the $\delta^N_{pol}$ term, which is tabulated in Table~\ref{tab:2} and is independent on the nuclear interaction. We use two potentials, one of phenomenological nature, the AV18~\cite{AV18} and one chiral interaction at next-to-next-to-next-to-leading order (N3LO)~\cite{entem2003}. Details on the expressions of the various terms can be found in Refs.~\cite{Hernandez2014,Nevo16,Ji2013}. As one can see from Fig.~\ref{fig:deut}, the potential dependence is quite small, of the order of 0.5$\%$. Rather than sampling potentials among those available in the literature, in future we aim at performing a statistical analysis of $\delta_{\rm TPE}$ by propagating the error bars associated to the parameters in the interaction through the observables themselves. This should enable us to investigate whether the above mentioned 2.5$\sigma$ deviation is originated from  the procedures used in nuclear physics or not. Work in this direction is in progress.

Similarly to the case of the Lamb shift, the hyperfine splitting energy is related to the magnetic radius $R_Z$ as
\begin{equation}
\label{eq:HFS}
\Delta E_{\rm hfs} = \delta^{\rm hfs}_{\rm QED}+\delta^{\rm hfs}_{\rm FS}(R_{Z})+\delta^{\rm hfs}_{\rm TPE}\,,
\end{equation}
where nuclear structure corrections come mostly from a TPE diagram.
Due to the fact that measurements of the hyperfine splitting are planned for $\mu$-D and $\mu$-$^3$He$^+$,
we are refining our tools to perform calculations of $\delta^{\rm hfs}_{\rm TPE}$ as well.
In case of the hyperfine splitting  $\delta^{\rm hfs}_{\rm TPE}$ is expected to be also related to magnetic properties of the nucleus~\cite{Pachucki,Friar,Chen_HFS}. To the purpose of enhancing our capabilities to compute magnetic properties, we investigate sum rules of the magnetic response function, starting from the deuteron. 
The magnetic response function is defined as
\begin{equation}
\label{resp}
 R(\omega)= \frac{1}{2J_0+1}\int \!\!\!\!\!\!\!\sum _{f} 
\left|\left\langle \Psi_{f} \left|\left| {\bm \mu} \right|\right| 
\Psi _{0}\right\rangle \right|
^{2}\delta\left(E_{f}-E_{0}-\omega \right)\,,
\end{equation}
where ${\bm \mu}$ is the magnetic dipole operator.  Here, $\left| \Psi _{0}\right\rangle$ and  $\left| \Psi _{f}\right\rangle $ denote the ground and excited states, respectively, while the sum/integral symbol is intended as a sum of discrete and an integral on continuum quantum numbers and states. The double bar denotes the reduced matrix element and the factor in front is an average on the projection of the ground state angular momentum $J_0$.

It is known that in nuclei magnetic transitions are not well described in impulse approximation, i.e., using one-body operators, and that two-body currents are important. Their expression has been derived in chiral effective field theory and their effect has been found to be very important in magnetic dipole moments and magnetic dipole transitions of light nuclei~\cite{Pastore13}. 
Here, we will develop our tools to accommodate the effect of leading order two-body currents from chiral effective field theory in the magnetic transitions of the deuteron.

\section{Two-body currents in the magnetic operator at next-to-leading order}

\begin{figure}[htb]
\centering
 \includegraphics*[width=5.cm]{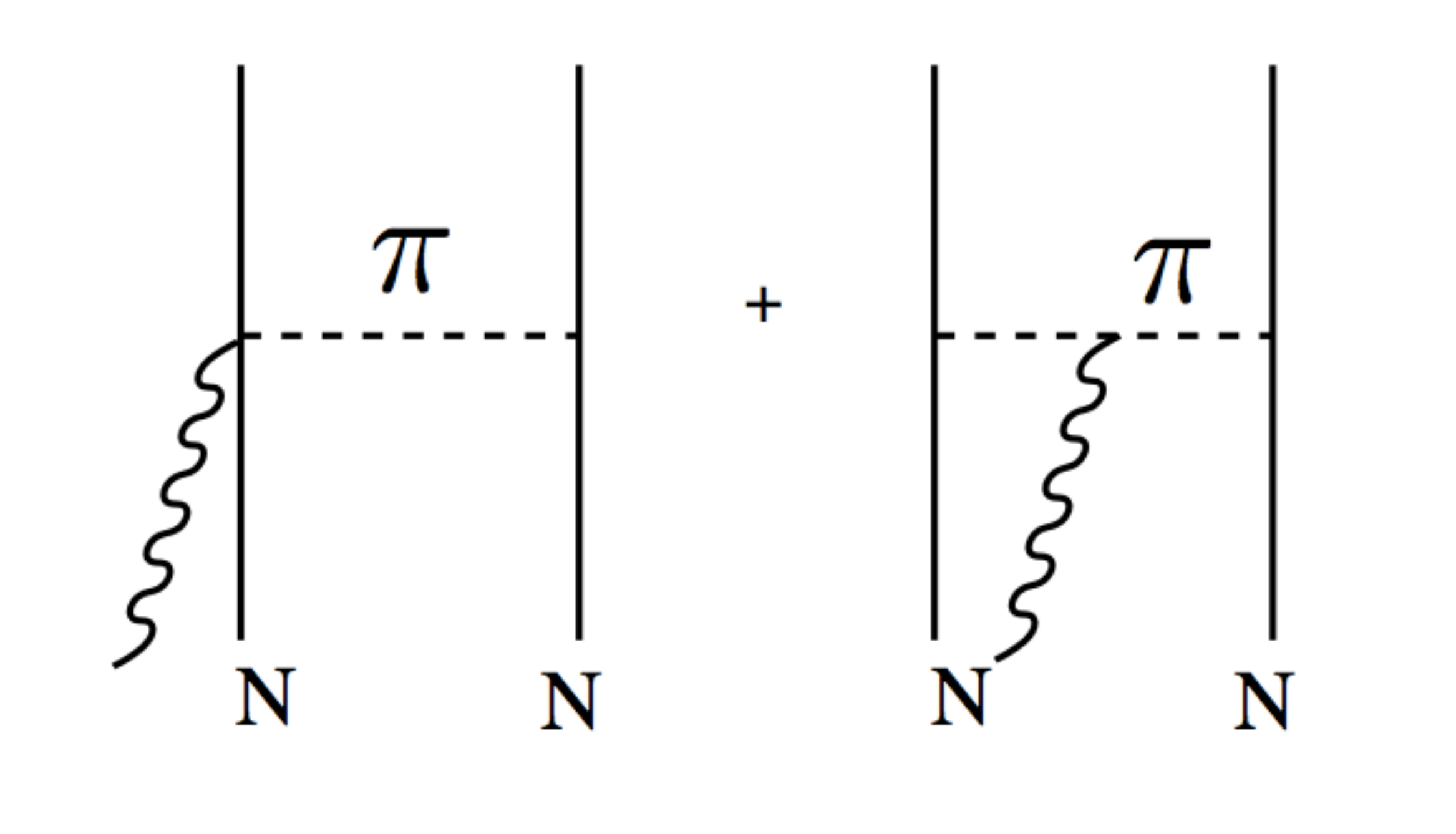}
\caption{Two-body currents in chiral effective field theory from a one-pion exchange diagram between the two nucleons: seagull (left) and pion in flight (right). The wiggle represents the electromagnetic interaction.}
\label{fig:mec} 
\end{figure}

In chiral effective field theory, similarly to what done for the strong force, the electromagnetic current can be expanded into many-body operators as 
\begin{equation}
\label{eq:j}
{\bf j}= \sum_i ~{\bf j}_i + \sum_{i<j}~ {\bf j}_{ij} +\dots \ .
\end{equation}
%Here,  ${\bf q}$ is the momentum associated with the external photon field.
Calculations performed using one-body operators only are named impulse approximation  calculations and 
are based on the idea that nuclear
properties are expressed as if the probing photon interacted only with individual nucleons. 
The impulse approximation corresponds to the leading (LO) order in chiral effective field theory. This description is improved  by accounting for the effects
of two-nucleon interactions onto the electromagnetic currents associated with nucleon
pairs. Two-body currents follow naturally
once a meson-exchange mechanism is invoked to describe the interactions among
nucleons. If one considers only the long range part of the nucleon-nucleon force, mediated by a one-pion exchange,
 two-body currents of one-pion nature emerge.
They result from photons hooking up with exchanged pions, are  shown in Fig.~\ref{fig:mec} 
by the seagull and pion-in-flight diagrams. 

The  one-body electromagnetic current operator in the non-relativistic limit consists of the usual
convection and spin-magnetization currents and in coordinate space read~\cite{pionnuclei}
\begin{eqnarray}
{\bf j}_i^c({\bf x})&=& \frac{e_i}{2m}\{{\bf p}_i,\delta({\bf x}-{\bf r}_i)\} ,\\
{\bf j}_i^s({\bf x})&=& i\frac{e\mu_i}{2m}{\bm \sigma}_i \times [{\bf p}_i,\delta({\bf x}-{\bf r}_i)]\,.
\end{eqnarray}
Here $ m$  is the nucleon mass (we keep the mass of the proton equal to the mass of the neutron) and $e_i$ and $\mu_i$ are the electric charge and magnetic moment of the nucleon, respectively, defined as
\begin{eqnarray}
e_i&=&\left( \frac{1+\tau^3_i}{2}\right)\\
\mu_i&=&\mu_p\left( \frac{1+\tau^3_i}{2}\right) + \mu_n \left( \frac{1-\tau^3_i}{2}\right)\,, 
\end{eqnarray}
 with $\tau_i^3$ being the third component of the nucleon isospin and $\mu_p=2.793$ and $\mu_n=-1.913$ in nucleon magneton $\mu_{N}$ units. 
Here,  nucleon coordinates and momenta are denoted by ${\bf r}_i$ and ${\bf p}_i$, respectively, while ${\bm \sigma_i}$ is the spin of the nucleon.

The one pion exchange  two-body currents of Fig.~\ref{fig:mec} appear at next-to-leading order (NLO) in chiral effective field theory and is the leading  two-body contribution.  The effect of NLO currents amounts to 70-80$\%$ of the total two-body currents contribution in magnetic properties of few-body systems~\cite{Piarulli}. 
We will call them ${\bf j}_{ij}^{\rm NLO}$ and separate them  into 
seagull current ${\bf j}_{ij}^{s}$ and the pion in-flight current ${\bf j}_{ij}^{\pi}$.  
Their expressions, more commonly found in momentum space, read~\cite{Piarulli}
\begin{eqnarray}
{\bf j}^{s}_{ij}({\bf k}_{i},{\bf k}_{j})&=-ie\frac{g^{2}_{A}}{F^{2}_{\pi}} G^{V}_{E}(q^{2})\left({\bm \tau}_{i} \times {\bm \tau}_{j} \right)^{3}\left( {\bm \sigma}_{i}\left( \frac{{\bm \sigma}_{j}\cdot {\bf k}_{j}}{\omega^{2}_{k_{j}}}\right)- {\bm \sigma}_{j}\left( \frac{{\bm \sigma}_{i}\cdot {\bf k}_{i}}{\omega^{2}_{k_{i}}}\right) \right)\\
{\bf j}^{\pi}_{ij}({\bf k}_{i},{\bf k}_{j})&=-ie \frac{g^{2}_{A}}{F^{2}_{\pi}} G^{V}_{E}(q^{2})\left({\bm \tau}_{i} \times {\bm \tau}_{j} \right)^{3} \left({\bf k}_{j}-{\bf k}_{i} \right) \left( \frac{{\bm \sigma}_{i}\cdot {\bf k}_{i}}{\omega^{2}_{k_{i}}} \right) \left( \frac{{\bm \sigma}_{j}\cdot {\bf k}_{j}}{\omega^{2}_{k_{j}}} \right)\,,\\
\end{eqnarray}
with 
\begin{equation}
{\bf j}_{ij}^{\rm NLO}({\bf k}_{i},{\bf k}_{j})={\bf j}^{s}_{ij}({\bf k}_{i},{\bf k}_{j}) +{\bf j}^{\pi}_{ij}({\bf k}_{i},{\bf k}_{j}) \,.
\end{equation}
Here, ${\bf k}_{i/j}$ is the momentum transferred to the nucleon $i$ or $j$,  $\omega_{k_{i/j}}^2=k_{i/j}^2+m_\pi^2$ is the
squared energy of the exchanged pion,
while ${\bm \tau}_{i/j}$ are nucleon isospin Pauli matrices. 
By performing the Fourier transform of these two-body currents, we obtain the expressions in coordinate space~\cite{Dubach01}
\begin{eqnarray}
\nonumber
{\bf j}^{s}_{ij}({\bf q})&= &-e\frac{m^{2}g^{2}_{A}}{4\pi F_{\pi}^{2}}G^{V}_E(q^{2})\left( {\bm \tau}_{i} \times {\bm \tau}_{j} \right)^{3}e^{i{\bf q}\cdot {\bf R}} \left[ e^{\frac{1}{2} i{\bf q}\cdot {\bf r}}{\bm \sigma}_{i}\left({\bm \sigma}_{j}\cdot \hat{r}\right)+ e^{-\frac{1}{2}i{\bf q}\cdot {\bf r}}{\bm \sigma}_{j}\left({\bm \sigma}_{i}\cdot \hat{r}\right)\right]\left( 1+\frac{1}{mr} \right)\frac{e^{-mr}}{mr}\\
\nonumber
{\bf j}^{\pi}_{ij}({\bf q})& =&  e\frac{2 g^{2}_{A}}{(2\pi)^{3}F^{2}_{\pi}} G^{V}_{E}(q^{2})\left({\bm \tau}_{i} \times {\bm \tau}_{j} \right)^{3} e^{i{\bf q}\cdot {\bf R}} \left( {\bm \sigma}_{i}\cdot \left(\frac{1}{2}{\bf q}-i{\bm \nabla}_{r} \right) \right)\!\!\!\left( {\bm \sigma}_{j}\cdot \left(\frac{1}{2}{\bm q}+i{\bm \nabla}_{r} \right) \right){\bm \nabla}_{r}I\left({\bf q},{\bf r} \right) \,,\\
\end{eqnarray}
where ${\bf q}$ is the momentum transfer and we use relative and center of mass coordinate of the two-interacting particles
\begin{eqnarray}
\nonumber
{\bf R} &= \frac{1}{2}\left({\bf r}_{i}+{\bf r}_{j} \right) \\
{\bf r} &= {\bf r}_{i}-{\bf r}_{j} \,.
\end{eqnarray}
In the current expression, the functions  $I({\bf q},{\bf r})$  arise when taking the Fourier transform of the pion in-flight term and are defined as
\begin{equation}
I({\bf q},{\bf r}) = \int d^{3}p \frac{e^{i{\bf p}\cdot {\bf r}}}{\left( m^{2}+\left( {\bf p}-\frac{1}{2}{\bf q} \right)^{2} \right)\left( m^{2}+\left( {\bf p}+\frac{1}{2}{\bf q} \right)^{2} \right)}\,.
\end{equation}

Given a current operator in coordinate space, 
the magnetic dipole operator is obtained from the latter using
\begin{equation}
{\bm \mu} = \frac{1}{2} \int d^{3}x \ {\bf x} \times {\bf j}({\bf x})\,.
\end{equation}
This general expression can be rewritten in the following way 
\begin{equation}
{\bm \mu} = \frac{1}{2}{\bf R} \times \int d^3x~  {\bf j}({\bf x}) + \frac{1}{2}\int d^{3}x ~({\bf x}-{\bf R})\times {\bf j}({\bf x})\,,
\end{equation}
and thus decomposed into two parts, 
where ${\bf R}$ is our center of mass coordinate. It is evident that the first term of the above equation will vanish if one considers an $A=2$ body problem in the center of mass frame. Indeed, since we will be studying the deuteron, we will only consider the second term. 
Since the ${\bf R}$-dependency in the current can be written as $e^{i{\bf q}\cdot{\bf R} }{\bf j}({\bf q},{\bf r})$,  the magnetic dipole operator obtained from the second term can be written by the curl of the  translational-invariant current operator at low ${\bf q}$ as~\cite{Pastore2008}
\begin{equation}
\label{curl}
{\bm \mu}({\bf r}) = \lim_{{\bf q} \rightarrow 0} -\frac{i}{2}{\bm \nabla}_{{\bf q}}\times {\bf j}({\bf q},{\bf r} )\,.
\end{equation}
Using the one-body current in Eq.~(\ref{curl}) one obtains the usual magnetic dipole operator as
\begin{equation}
{\bm \mu}^{\rm LO}_i= \mu_N \left[ \left( \frac{\mu^S+\mu^V\tau^3_i}{2}\right){\bm \sigma}_i +\left( \frac{1+\tau^3_i}{2}\right)  {\bm \ell }_i \right ] \,,
\end{equation}
where $\mu^{S/V}$ are the isoscalar and isovector nucleon magnetic moments, $4.7$ and $0.88$ in nucleon nucleon magneton $\mu_N$ units, respectively. This one-body operator is the leading order term in chiral effective field theory.
To obtain the two-body corrections to the above one-body operator, we can 
plug in the expression of the seagull and pion in flight currents in Eq.~(\ref{curl}), to obtain  the magnetic dipole operator due to the seagull and pion in flight diagrams, respectively, as
\begin{eqnarray}
{\bm \mu}^{s}_{ij} &=& -e\frac{m g^{2}_{A}}{16 \pi F^{2}_{\pi}}\left( {\bm \tau}_{i} \times {\bm \tau}_{j} \right)^{3}\left[\hat{\bf r}(\hat{\bf r}\cdot \left({\bm \sigma}_{i}\times {\bm \sigma}_{j} \right))-{\bm \sigma}_{i}\times {\bm \sigma}_{j} \right] f(r)\\
\nonumber
{\bm \mu}^{\pi}_{ij}&=& -\frac{eg^{2}_{A}m}{16\pi F^{2}_{\pi}}\left( {\bm \tau}_{i} \times {\bm \tau}_{j} \right)^{3}\left[(\hat{\bf r} \cdot {\bm \sigma}_{j})(\hat{\bf r}\times {\bm \sigma}_{i})-(\hat{\bf r} \cdot {\bm \sigma}_{i})(\hat{\bf r}\times {\bm \sigma}_{j}) \right] f(r) \\
&-&\frac{eg^{2}_{A}m}{8\pi F^{2}_{\pi}}\left( {\bm \tau}_{i} \times {\bm \tau}_{j} \right)^{3} \left({\bm \sigma}_{i}\times {\bm \sigma}_{j} \right) Y(r)\,.
\end{eqnarray}
Here, the functions $f(r)$ and $Y(r)$, with $r=|{\bf r}|$ are
\begin{eqnarray}
\nonumber
f(r) & =& \left( 1+\frac{1}{mr} \right) {e^{-mr}}\,, \\
Y(r) & = &  \frac{e^{-mr}}{mr}\,.
\end{eqnarray}
Thus, at next-to-leading-order the two-body magnetic moment is given by the sum ${\bm \mu}^{\rm NLO}_{ij} = {\bm \mu}^{s}_{ij}+{\bm \mu}^{\pi}_{ij}$ as~\cite{pionnuclei}
\begin{equation}
{\bm \mu}^{\rm NLO}_{ij} = -\frac{eg^{2}_{A}m}{8\pi F^{2}_{\pi}}\left({\bm \tau}_{i}\times {\bm \tau}_{j} \right)^{3}\left[\left(1+ \frac{1}{mr} \right)\left(\left({\bm \sigma}_{i}\times {\bm \sigma}_{j} \right)\cdot \hat{\bf r} \right)\hat{\bf r} -\left({\bm \sigma}_{i}\times {\bm \sigma}_{j} \right) \right]e^{-mr}\,.
\end{equation}

Finally, the magnetic dipole operator will be given by a leading-order one body component and a next-to-leading order two-body component as
\begin{equation}
{\bm \mu}=\sum_i {\bm \mu}^{\rm LO}_i + \sum_{i<j} {\bm \mu}^{\rm NLO}_{ij}\,.
\end{equation}

Next, we will implement these operators in our deuteron calculation of magnetic properties. It is important to remember that, in a many-body nucleus with $A>2$, the NLO two-body correction contains an other term which explicitly depends on ${\bf R}$ even at small ${\bf q}$ and is called Sachs term~\cite{pionnuclei}. Furthermore, as already mentioned, other corrections exist at higher order in chiral effective field theory and have been accounted for, e.g., in Refs.~\cite{Pastore13,Piarulli}. From those calculations,  it is evident that the NLO two-body currents accounts for up to 70-80 $\%$ of the total two-body current effects.

%The explicit connection between many-body potentials and
%many-body current operators is provided by the continuity equation
%imposed by the gauge invariance of the theory:
%\begin{equation}
%{\bf q}\cdot {\bf j}=\left[H ,\, \rho \,\right ] \ ,
%\label{eq:continuity}
%\end{equation}
%where $H$ is the nuclear Hamiltonian, whose many-body operatorial decomposition is given in
%Eq.~(\ref{eq:hamiltonian}), and $[\dots,\dots]$ denotes a commutator. 

\section{Results}

Using the above expressions for the one- and two-body magnetic dipole operators, we now  study some magnetic observables in the deuteron. First of all, due to the fact that ${\bm \mu}^{\rm NLO}_{ij}$ is of isovector nature, contributions of two-body currents at next-to-leading order will vanish in the magnetic moment of the deuteron. Thus, we concentrate on break-up observables, such as the sum rules of the magnetic dipole transition function.
In the following we will investigate quantities of this kind
\begin{equation}
\label{sumrule}
m_n = \int d\omega R(\omega) \omega^{n}\,
\end{equation} 
with $n=-1$ and $0$, and $R(\omega)$ as in Eq.~(\ref{resp}). In particular, for the case of $n=-1$ this quantity is related to the magnetic susceptibility, and is the magnetic analogous of the electric dipole polarizability.
Such sum rules have also been calculated in the past, see e.g. Ref.~\cite{Arenhovel}, so we can compare our calculations with similar theoretical calculations. A comparison with experiment is more difficult due to the fact that in sum rules one has to integrate the strength up to infinity  and one has to clearly separate out the contribution from other multipoles. 

We perform our analysis by solving the deuteron via a diagonalization of the intrinsic Hamiltonian on the harmonic oscillator basis. We can perform calculations with any realistic two-body potential and will show here results with either the AV18~\cite{AV18} or the N3LO chiral potential~\cite{entem2003}.
 The calculation of the sum rules follows according to Ref.~\cite{Hernandez2014}. 

First, to check our numerical ability to calculate magnetic sum rules we compare our LO calculations, corresponding to the use of a one-body operator only ${\bm \mu}=\sum_i {\bm \mu}^{\rm LO}_i$,
 with results by Arenh\"{o}vel~\cite{Arenhovel,Arenhovel_private}. In Ref.~\cite{Arenhovel}, results were obtained with the Bonn r-space potential, but here we present a comparison with a more modern interaction, the AV18~\cite{AV18} potential.
 As one can see in Table~\ref{tab:3} we obtain a rather good agreement. The small sub-percentage difference  has to be attributed to the fact that we have integrated the magnetic dipole strength obtained from Ref.~\cite{Arenhovel_private} using Eq.~(\ref{sumrule}), while in our case we computed the sum rule directly as an expectation value on the ground-state. 
%The difference is larger for the case of $m_1$, where one does need to integrate up to high energy to saturate the sum rule. We have integrated up to 500 MeV.
To confirm our numbers, we have performed two independent implementations  and obtained very nice numerical agreement among them, at the level of 0.1$\%$ or better. 

\begin{table}[htb]

\centering

\caption{Sum rules of the magnetic response function of the deuteron, calculated with the AV18 potential~\cite{AV18}, using a one-body magnetic dipole operator.}

\label{tab:3}

\begin{tabular}{l|l|l}

\hline\noalign{\smallskip}

&$m_{-1}$& $m_0$\\% & $m_1$\\

\noalign{\smallskip}\hline\noalign{\smallskip}

This work & 13.9 fm$^3$ &  0.245 fm$^2$ \\% & 0.0137 fm \\

Ref.~\cite{Arenhovel_private} & 14.0 fm$^3$ &  0.244 fm$^2$ \\% & 0.0126 fm\\

\noalign{\smallskip}\hline

\end{tabular}

\end{table}

Next, we introduce the two-body correction to the magnetic dipole operator at NLO and compare to the LO calculation in Table~\ref{tab:4}. In this case we will use a potential from chiral effective field theory at N3LO~\cite{entem2003}. It has to be noted that, from the chiral effective field theory stand point, such calculations are not fully consistent, since potential and currents are not taken at the same order, but at this point our objective is to prepare our tools for a more sophisticated calculation to be carried out in the future.

\begin{table}[htb]

\centering

\caption{Sum rules of the magnetic response function of the deuteron, calculated with a chiral interaction at N3LO~\cite{entem2003}  and a magnetic dipole operator at LO and at NLO.
}

\label{tab:4}

\begin{tabular}{l|l|l}

\hline\noalign{\smallskip}

&$m_{-1}$& $m_0$ \\%& $m_1$\\

\noalign{\smallskip}\hline\noalign{\smallskip}

LO &      14.0 fm$^3$ & 0.245 fm$^2$ \\% & 0.0106 fm \\

LO+ NLO & 15.1 fm$^3$ & 0.277 fm$^2$ \\% & 0.0184 fm\\

\noalign{\smallskip}\hline

\end{tabular}

\end{table}

%*****

Given that, to the best of our knowledge, no calculation with just the NLO two-body current is available in the literature, we have compared our numerics against an independent computation of the multipole matrix elements of tensor currents~\cite{Wendt} and found a numerical agreement at the $0.1\%$ level or better.
Overall, we find the effect of two-body currents to be between $5$ and $11\%$ depending on the order of the sum rule. Notably, the effect is bigger on the $m_0$ sum rule than on the $m_{-1}$. This fact indicats  that two-body currents have more of an effect at larger energies than at lower energies. This is consistent with results in the literature~\cite{Arenhovel} from phenomenological currents and potentials. 
Clearly, even if their effect might be as small as a $5\%$, two-body currents need to be taken into account when doing precision physics as in the studies of nuclear structure corrections in muonic atoms.

 In the case of the Lamb shift, magnetic contributions to the $\delta_{\rm TPE}$ diagram appear via  $\delta_{M}$. This term is very small, amounting to $0.007$ meV in the deuteron when the LO magnetic operator is used.  With the addition of the two-body contributions at NLO, its contribution goes from 0.007 to $0.009$ meV, with a 20$\%$ enhancement. While the overall  contribution of two-body magnetic currents is very small in the Lamb shift due to the fact the $\delta_M$ itself is small, it is expected to be larger in the hyperfine splitting, where the magnetic current distributions play a role, see Ref.~\cite{Pachucki,Friar,Chen_HFS}.

\section{Conclusion}

We have reviewed the status of the nuclear structure calculations performed by the TRIUMF - Hebrew University group for the Lamb shift in muonic atoms and  presented new calculations of magnetic sum rules in the deuterium with two-body currents. We find an effect of 5 and 11$\%$ on the $m_{-1}$ and $m_0$ magnetic sum rules, respectively, which is consistent with previous investigations. Two-body currents are expected to provide a non-negligible contribution in nuclear structure corrections to the hyperfine splitting in muonic atoms. While the presented results constitute a necessary ingredient for a detailed study of the hyperfine splitting corrections, a complete analysis is left for future work.

\section{Acknowledgments}

 We thank Saori Pastore and the other members of the TRIUMF-Hebrew University collaboration for helpful discussions. We are indebted to Hartmuth Arenh\"{o}vel for providing us with the magnetic dipole strength of the deuteron that served as an important check.
  This work was supported in parts by the Natural Sciences
  and Engineering Research Council (NSERC), the National Research
  Council of Canada. This work was performed under the auspices of the U.S. Department of Energy by Lawrence Livermore National Laboratory under Contract DE-AC52-07NA27344. Funding was provided by LLNL, Lawrence Fellowship Program.


\begin{thebibliography}{99}

\bibitem{Pohl:2010zza} R.~Pohl {\it et al.},
%  title = {The size of the proton},
Nature {\bf 466}, 213 (2010).


\bibitem{Antognini:1900ns} A. Antognini {\it et al.}, 
Science {\bf 339}, {417} (2013).

\bibitem{science2016} R.~Pohl {\it et al.}, Science {\bf 353}, 669-673 (2016).


\bibitem{Krauth_paper} J.J. Krauth {\it et al.}, Annals Phys. {\bf 366} 168-196 (2016). 


\bibitem{Franke} B.~Franke {\it et al.}, 1705.00352.


\bibitem{Hernandez2014} O.J. Hernandez and C. Ji, S. Bacca, N.~Nevo-Dinur and N.~Barnea,
Phys. Lett. B {\bf 736}, 344 - 349 (2014).



\bibitem{Nevo16}N.~Nevo Dinur, C.~Ji, S.~Bacca,   and N.~Barnea,  Physics Letters B {\bf 755}, 380-386 (2016). 


\bibitem{Ji2013}
%%Nuclear polarizability corrections in the $\mu-^4$He$^+$ Lamb shift
C.~Ji, N.~Nevo Dinur, S.~Bacca, N.~Barnea, Phys. Rev. Lett. {\bf 111}, 143402 (2013).

\bibitem{Nevo2014}  N.~Nevo Dinur, N.~Barnea, S.~Bacca, C.~Ji, Phys. Rev. C {\bf 89}, 064317 (2014).
%\bibitem{Machleidt11} R. Machleidt and D.R. Entem, 
%%	title = "Chiral effective field theory and nuclear forces ",
%Physics Reports {\bf 503} 1-75 (2011).
\bibitem{entem2003} D.R.~Entem and R.~Machleidt, Phys. Rev. C {\bf 68}, 041001 (2003).

\bibitem{Epelbaum09} E. Epelbaum, H.W.~Hammer, U.-G.~Mei\ss{}ner, Rev. Mod. Phys. {\bf 81}, 1773--1825 (2009).


\bibitem{AV18} R.B.~Wiringa, {\it et al.}, Phys. Rev. C {\bf 51}, 38 (1995).


\bibitem{Javier2016} O.~J.~Hernandez, N.~Nevo Dinur, C.~Ji, S.~Bacca, and N.~Barnea,
Hyperfine Interact {\bf 237}, 158 (2016).

\bibitem{Pachucki11} K.~Pachucki, Phys. Rev. Lett. {\bf 106} 193007 (2011).

\bibitem{Friar13} J.L.~Friar, Phys. Rev. C {\bf 88}, 034003  (2013).

\bibitem{Carlsson14}  C.E.~Carlsson, M.~Gorchtein and M.~Vanderhaeghen, Phys. Rev. A {\bf 89}, 022504 (2014).

\bibitem{Pachucki15}   K.~Pachucki and A.~Wienczek, Phys. Rev. A {\bf 91}, 040503(R) (2015).



\bibitem{Mohr:2012tt} 
  P.~J.~Mohr, B.~N.~Taylor and D.~B.~Newell,
  %``CODATA Recommended Values of the Fundamental Physical Constants: 2010,''
  Rev.\ Mod.\ Phys.\  {\bf 84}, 1527 (2012).

\bibitem{deut_spect} R.~Pohl {\it et al.}, Metrologia {\bf 54}, L1 (2017).



\bibitem{Sick} I.~Sick and D.~Trautmann, Nucl. Phys. A {\bf 637}, 559-575 (1998).



%\bibitem{Parthey} C.G.~Parthey {\it et al.}, Phys. Rev. Lett. {\bf 104}, 233001 (2010).




%\bibitem{Epelbaum12}  E.~Epelbaum and U.-G.~Mei\ss{}ner,
%%      title          = "{Chiral dynamics of few- and many-nucleon systems}",
%Ann.~Rev.~Nucl.~Part.~Sci. {\bf 62}, 159-185 (2012).
\bibitem{Pachucki} K.~Pachucki,
%%  title = {Nuclear vector polarizability correction to hyperfine splitting},
 Phys. Rev. A  {\bf 76}, 022508 (2007).

\bibitem{Friar} J.L. Friar and G.L. Payne,
%title = "The nuclear physics of hyperfine structure in hydrogenic atoms ",
Phys. Lett. B {\bf 618}, 68 - 76 (2005).

\bibitem{Chen_HFS} C.~Ji, O.J.~Hernandez, N.~Nevo Dinur, S.~Bacca and N.~Barnea, in preparation.

\bibitem{Pastore13} S.~Pastore, S.C.~Pieper, R.~Schiavilla and R.B.~Wiringa,  
%  title = {Quantum Monte Carlo calculations of electromagnetic moments and transitions in $A\le9$ nuclei with meson-exchange currents derived from chiral effective field theory}
 Phys. Rev. C {\bf 87} 035503 (2013).
\bibitem{pionnuclei} T.~Ericson and W.~Weise, ``Pion s and Nuclei'', Clarendon Press, Oxford (1988).

\bibitem{Piarulli} M.~Piarulli, L.~Girlanda, L.E.~Marcucci, S.~Pastore, R.~Schiavilla and M.~Viviani, 
Phys. Rev. C {\bf 87} 014006 (2013).

\bibitem{Dubach01} J. Dubach, J. H. Koch, and T.W.~Donnelly, Nucl. Phys. A. {\bf 271} 279-316 (1976).


\bibitem{Pastore2008} S.~Pastore, R.~Schiavilla and J.L.~Goity, Phys. Rev. C {\bf 78}, 064002 (2008).



\bibitem{Arenhovel} H.~Arenh\"{o}vel and M.~Sanzone, Few-Body Systems, Supplement {\bf 3}, 1-183 (1991).


\bibitem{Arenhovel_private} H.~Arenh\"{o}vel, private communication (2016).

\bibitem{Wendt} K.A.~Wendt {\it et al.}, in preparation. 







\end{thebibliography}
\end{document}